\documentclass[a4paper]{jpconf}
\usepackage{iopams}  

\usepackage{graphicx,hyperref}

\begin{document}

\title{MHD Memes}

\author{R~L Dewar,$^{1,2}$ R Mills$^1$ and M~J Hole$^1$}

\address{$^1$Research School of Physical Sciences and Engineering, The Australian National University,
Canberra ACT 0200, Australia\\
$^2$Graduate School of Frontier Sciences, University of Tokyo, Kashiwa-City
Kashiwanoha 5-1-5 Chiba 277-8561,
Japan}

\ead{robert.dewar@anu.edu.au}

\begin{abstract}
The celebration of Allan Kaufman's 80th birthday was an occasion to reflect on a career that has stimulated the mutual exchange of ideas (or memes in the terminology of Richard Dawkins) between many researchers. This paper will revisit a meme Allan encountered in his early career in magnetohydrodynamics, the continuation of a magnetohydrodynamic mode through a singularity, and will also mention other problems where Allan's work has had a powerful cross-fertilizing effect in plasma physics and other areas of physics and mathematics.

\end{abstract}


  \newcommand{\Fig}[1]{figure~\ref{fig:#1}}

 \newcommand{\Eqn}[1]{(\ref{eq:#1})}

 \newcommand{\Eqns}[2]{(\ref{eq:#1}--\ref{eq:#2})}

 \newcommand{\Sec}[1]{Sec.~\ref{sec:#1}}
 \newcommand{\sgn}{{\rm sgn}\, }
 \renewcommand{\Re}{{\rm Re}\, }
 \renewcommand{\Im}{{\rm Im}\, }

 \newcommand{\iotabar}{\mbox{$\,\iota\!\!$-}}

\renewcommand{\vec}[1]{\mbox{$\bf #1$}}
 \newcommand{\xiv}{\mbox{\boldmath$\xi$}}
 \newcommand{\tauv}{\mbox{\boldmath$\tau$}}
 \newcommand{\Id}{\mbox{\boldmath\sf I}}

\newcommand{\const}{\mbox{const}}
\newcommand{\half}{ \frac{\mbox{\small{1}}}{\mbox{\small{2}}} }

\newcommand{\esub}[1]{ {\bf e}_{#1} }
\newcommand{\esup}[1]{ {\bf e}^{#1} }
\newcommand{\dotv}{  \mbox{\boldmath\(\cdot\)} }
\newcommand{\ddotv}{  \mbox{\boldmath\(:\)} }
\newcommand{\cross}{  \mbox{\boldmath\(\times\)} }
\newcommand{\dsub}[1]{ \partial_{#1} }
\renewcommand{\d}{ {\rm d} } 

\newcommand{\grad}{  \mbox{\boldmath\(\nabla\)} }
\newcommand{\divv}{  \mbox{\boldmath\(\nabla\cdot\)} }
\newcommand{\curl}{  \mbox{\boldmath\(\nabla\times\)} }
 \newcommand{\jump}[1]{\left[ \!\left[ #1 \right] \! \right]} 

 \newcommand{\curlz}{\nabla_0 \times}
 \newcommand{\divz}{\nabla_0 \cdot}
 \newcommand{\deriv}[2][\null]{\frac{d#1}{d#2}}
 \newcommand{\derivn}[3][\null]{\frac{d^#3 #1}{d#2^#3}}
 \newcommand{\dderiv}[3][\null]{\frac{d^2#1}{d#2 d#3}}
 \newcommand{\pderiv}[2][\null]{\frac{\partial #1}{\partial #2}}
 \newcommand{\pderivn}[3][\null]{\frac{\partial^#3 #1}{\partial #2^#3}}
 \newcommand{\ppderiv}[3][\null]{\frac{\partial^2 #1}{\partial #2 \partial #3}}
 \newcommand{\pvarderiv}[2][\null]{\frac{\delta #1}{\delta #2}}
 \newcommand{\pvarderivn}[3][\null]{\frac{\delta^#3 #1}{\delta #2^#3}}
 \newcommand{\ppvarderiv}[3][\null]{\frac{\delta^2 #1}{\delta #2 \delta #3}}
 \newcommand{\cosb}[1]{\cos \left(#1\right)}
 \newcommand{\sinb}[1]{\sin \left(#1\right)}
 \newcommand{\tanb}[1]{\tan \left(#1\right)}
 \newcommand{\inprod}[2]{\left\langle #1 , #2 \right\rangle}

 \newcommand{\followupnote}[1]{\begin{center}{FOLLOWUP:\qquad\framebox[0.7 \textwidth]{\framebox[0.68 \textwidth]{\parbox{0.66 \textwidth}{#1}}}}\end{center}}
 \renewcommand{\output}[1]{\begin{center}{OUTPUT\qquad:\parbox{0.66 \textwidth}{\texttt{#1}}}\end{center}}
 \newcommand{\code}[1]{\begin{center}{CODE:\qquad\parbox{0.66 \textwidth}{\texttt{#1}}}\end{center}}

\section{Introduction}
	\label{sec:Intro}

Richard Dawkins \cite{Dawkins_76,Dawkins_86}, in discussing the evolution of ideas, introduced the concept of  \emph{memes} in the following terms ``\dots\  new replicators are not DNA \ldots\  patterns of information that can thrive only in brains \ldots\  or books, computers and so on \ldots\  called memes to distinguish them from genes \ldots\  passed from brain to brain \ldots\  As they propagate they can change Ñ mutate \ldots\  memic evolution \ldots'' 
	
Applying and extending this biological analogy to the evolution of science, we can  regard scientific papers as the ``organisms'' in which multiple memes are expressed, and say that cross-fertilization leads to hybrid vigour.
	
One of Allan Kaufman's greatest contributions to plasma physics has been his role in adapting and developing powerful theoretical and mathematical methods, applying them to plasma physics problems, and propagating these ideas to following generations through his ability as a great teacher.
	
My (RLD) research interests have intersected, and continue to intersect, with Allan's varied research interests. In this paper I have chosen to illustrate this with a few examples from magnetohydrodynamics (MHD). This may seem a little unexpected, because Allan is not usually thought of as an MHD theorist, but in section \ref{sec:Newcomb} I revisit a problem he was familiar with in his early career, namely the continuation (or lack thereof) of solutions of the Newcomb equation for linearized ideal marginal MHD modes. The disconnection between solutions on either side of a mode rational surface is a tricky point to explain to newcomers in the field and I present a new approach which I hope makes this phenomenon clearer.
	
In section \ref{sec:OtherMHD} I mention some other MHD problems I have encountered which have intersections with Allan's research.

\section{The connection problem at a singular point of the Newcomb equation}
	\label{sec:Newcomb}

It is probably not widely known that some of Allan's earliest work was on magnetohydrodynamic (MHD) stability theory \cite{Newcomb_Kaufman_1961}. Newcomb's famous paper on ideal-MHD stability theory in cylindrical geometry \cite{Newcomb_60} acknowledges Allan's critical reading of the manuscript while his Ref. 24 is the note ``M. N. ROSENBLUTH (private communication through A. N. Kaufman),'' showing that Allan's cross-pollinating role in plasma physics started very early on.

It happens that these papers are highly relevant to our current research \cite{Hudson_Hole_Dewar_07,Hole_Hudson_Dewar_07,Dewar_etal_08} on equilibrium and stability in a multiple-region relaxed-MHD plasma model. This is motivated by the desire to construct a well-posed three-dimensional MHD equilibrium theory, based on Taylor relaxation in regions separated by arbitrarily thin ideal-MHD toroidal surfaces that act as barriers to field-line chaos. Not only does this involve the Hamiltonian nonlinear dynamics of the field-line flow, but the Hamiltonian of the field lines must itself be determined self-consistently using MHD theory. 

To understand the nature of these ideal-MHD barriers better we have recently \cite{Mills_07,Mills_Hole_Dewar_09} returned to the cylindrical limit and looked at the problem of whether the assumed barrier between two relaxed regions of different pressure can be constructed as the zero-width limit of a finite-width ideal-MHD region with a suitably chosen physical pressure profile. [The criterion for physicality of the pressure profile is that  the pressure $P(r)$ must be non-negative, where $r$ is the distance from the $z$-axis.] This problem was first studied by Newcomb and Kaufman \cite{Newcomb_Kaufman_1961,Newcomb_60}.

The ``Newcomb equation'' \cite{Newcomb_60} is satisfied at marginal stability (i.e. when the growth rate $\gamma$ is zero) by $\xi(r) \equiv \xiv\dotv\esub{r}$, where $\xiv \propto \exp i(m\theta + kz)$ is the plasma displacement away from equilibrium, $\theta$ being the angle about the $z$-axis and $\esub{r}$  the unit vector in the radial direction:
\begin{equation}
	\frac{d}{dr}\left(f\, \frac{d\xi}{dr}\right)-g\,\xi = 0,
		\label{eq:NewcombEL}
\end{equation}
where
\begin{eqnarray}
	f(r) &\equiv& \frac{r[m B_{\theta}(r)+k r B_z(r)]^2}{k^2r^2+m^2},
		\label{eq:Newcombf}\\
	g(r) &\equiv& \frac{2k^2r^2}{k^2r^2+m^2}\frac{dP}{dr}
	+\frac{[mB_{\theta}(r)+k r B_z(r)]^2(k^2r^2+m^2-1)}	{r(k^2r^2+m^2)}\nonumber\\
	&&+\frac{2k^2r[k^2r^2B_z(r)^2-m^2 B_{\theta}(r)^2]}{(k^2r^2+m^2)^2}.
		\label{eq:Newcombg}
\end{eqnarray}

For given integers $m$ and $n \equiv -Rk \neq 0$,  a \emph{mode rational surface} is one where $q(r_s) = m/n$, with $q(r) \equiv rB_z/R B_{\theta}$ and $R$ being the periodicity length of the cylinder. At such a rational surface $r_s$, \Eqn{NewcombEL} has a \emph{singular point} because the coefficient of the highest derivative, $f(r)$,  vanishes there.  Defining $x \equiv r - r_s$, the Taylor expansions of $f$ and $g$ about $r_s$ are of the form $f(r) = f''(r_s)x^2/2 + O(x^3)$ and $g(r) = g(r_s) + O(x)$, with $f''(r_s) > 0$ provided $q'(r_s) \neq 0$ and $g(r_s) \neq 0$ provided $P'(r_s) \neq 0$.  The general solution of \Eqn{NewcombEL} on intervals not including $r_s$ is found as a linear superposition of two functions whose Frobenius expansions (see e.g. \cite{Ince_1956,Dewar_Persson_93}) begin with the fractional powers $|x|^{-1/2 \pm \mu}$, where
\begin{equation}
   	\mu \equiv \frac{\sqrt{1+8g(r_s)/f''(r_s)}}{2}.
	\label{eq:mudef}
\end{equation}

We assume $\mu$ is real, otherwise the system is interchange-unstable by the Suydam criterion. The solutions with leading terms $|x|^{-1/2-\mu}$ and $|x|^{-1/2+\mu}$ were called by Newcomb the \emph{large} and \emph{small} solutions, respectively, the large solution being dominant as $x \rightarrow 0$.

In this paper we discuss a point raised by Newcomb \cite[p. 241]{Newcomb_60}: ``In general it is not possible to continue an Euler--Lagrange solution past a singular point.'' To this he adds the footnote: ``This is true only because $\xi$ is a real variable. If it were complex we could go around the singular point by analytic continuation, but the resulting solutions would generally be multivalued.'' The issue is interesting in the context of the present paper because this meme is perhaps expressed in a mutated form in Allan's later papers on mode conversion (e.g. \cite{Kaufman_Friedland_87}), which also involves continuation through singularities.

The natural framework in which to understand the singular solutions of the Newcomb equation is generalized function theory because the solutions are not defined pointwise (being undefined at $r = r_s$). In our previous work \cite{Manickam_Grimm_Dewar_81,Grimm_Dewar_Manickam_83,Dewar_Pletzer_90}, using the singular solutions of the Newcomb equation or the two-dimensional generalization of it due to Bineau \cite{Bineau_1962}, we have followed the approach and notation of of Gel'fand and Shilov  \cite{Gelfand_Shilov_1964} based on inner products with all members of a space of smooth test functions.

However this approach is rather abstract and difficult to visualize, so in the current paper we present an alternative approach, where the required generalized functions are defined as limits of sequences, parametrized by $\delta$,  of smooth functions that tend to weak solutions of the Newcomb equation as $\delta \rightarrow +0$. This is in the spirit of Lighthill's \cite{Lighthill_1958} approach to generalized function theory.

To do this we study a model Newcomb equation, in which $f(x) = x^2$ and $g(x) = \mu^2 - 1/4$. The weak solution space is spanned by the \emph{four} Gel'fand-Shilov generalized functions
\begin{eqnarray}
	\Xi^{\rm Fr}_{-}(x) & \equiv & |x|^{-1/2-\mu}\,\sgn x \nonumber \\
	\Xi^{\rm Fr}_{+}(x) & \equiv & |x|^{-1/2-\mu} \nonumber \\
	\xi^{\rm Fr}_{-}(x) & \equiv & |x|^{-1/2+\mu}\,\sgn x \nonumber \\
	\xi^{\rm Fr}_{+}(x) & \equiv & |x|^{-1/2+\mu} , \label{eq:unregParity} 
\end{eqnarray}
where $\sgn x$ denotes the sign of $x$. On the left-hand sides we have used the notation of Dewar and Persson \cite{Dewar_Persson_93} in which the large and small Frobenius solutions are indicated by using upper and lower case greek, respectively; odd and even parity solutions are indicated by subscript $-$ and $+$, respectively.

Alternatively, we may use the ``single-sided'' generalized functions $|x|_{\pm}^{\lambda} \equiv \frac{1}{2}(|x|^{\lambda}  \pm |x|^{\lambda}\,\sgn x)$ whose support is the positive ($x > 0$) or negative ($x < 0$) half line. When restricted to their domains of support these single-sided solutions correspond to the classical pointwise solutions Newcomb meant when he said solutions of the Newcomb equation could not be continued through the singular point. As the domains of support are to the left (L) and right (R) of the origin, we denote the corresponding generalized function solutions of the Newcomb by subscripts L and R
\begin{eqnarray}
	\Xi^{\rm Fr}_{\rm L}(x) & \equiv & |x|_{-}^{-1/2-\mu} \nonumber \\
	\Xi^{\rm Fr}_{\rm R}(x) & \equiv & |x|_{+}^{-1/2-\mu} \nonumber \\
	\xi^{\rm Fr}_{\rm L}(x) & \equiv & |x|_{-}^{-1/2+\mu} \nonumber \\
	\xi^{\rm Fr}_{\rm R}(x) & \equiv & |x|_{+}^{-1/2+\mu} . \label{eq:unregLR} 
\end{eqnarray}

In ideal MHD the large solutions are rejected as they are not square integrable (hence give infinite kinetic energy), thus reducing the dimensionality of the solution space to 2 as expected for second-order differential equations. (However, the large solutions are essential in resistive stability theory for matching asymptotically to the resistive internal layer in the neighbourhood of the singular point.)

To apply the Lighthill approach we use a regularization appropriate to ideal-MHD stability studies, in which the only regularizing effect is that of inertia from the mass density $\rho$. This comes into play when the growth rate $\gamma$ is nonzero, adding to $f$ a positive term proportional to $\rho\gamma^2$ (see e.g. \cite{Dewar_etal_04}) and thus removing the singularity but approximating the Newcomb equation arbitrarily closely as $\gamma \rightarrow 0$. Thus we study the \emph{regularized model Newcomb equation}
\begin{equation}
	\frac{d}{dx}(x^2 + \delta^2)\frac{d}{dx}\xi - \left(\mu^2 - \frac{1}{4}\right)\xi = 0,
		\label{eq:NewcombReg}
\end{equation}
whose general solution is a linear combination of  the respectively odd and even functions $\Im P_{\mu - 1/2}(ix/\delta)$ and $\Re P_{\mu - 1/2}(ix/\delta)$, where $P_{\nu}(z) \equiv P^0_{\nu}(z)$ denotes a Legendre function of the first kind  \cite[chapter 8]{Abramowitz_Stegun_72}.

\begin{figure}[htbp]
	\centering
	\begin{tabular}{cc}
	\includegraphics[width=7cm]{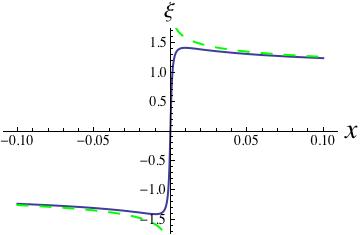}   &  \includegraphics[width=7cm]{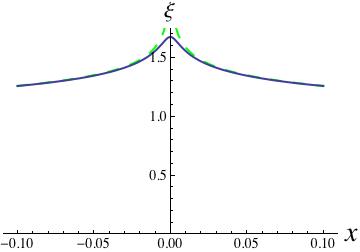}
	\end{tabular}
	\caption{Left panel:  regularized odd solution $\xi^{\rm Fr}_{-}(x|0.001)$ (solid line) and  unregularized odd solution $\xi^{\rm Fr}_{-}(x)$ (dashed line). Right panel: regularized even solution $\xi^{\rm Fr}_{+}(x| 0.005)$ (solid line) and unregularized even solution $\xi^{\rm Fr}_{+}(x)$ (dashed line). In both cases $\mu = 0.4$.}
	\label{fig:OddEven}
\end{figure}

By expanding $P_{\mu - 1/2}(ix/\delta)$ in powers of $\delta$ and choosing the superposition coefficients appropriately we can now define Lighthill approximating functions for the odd and even small solutions
\begin{eqnarray}
	\xi^{\rm Fr}_{-}(x|\delta) & \equiv &
	\frac{(2\delta)^{\mu -\frac{1}{2}}\Gamma \left(\mu +\frac{1}{2}\right)^2 }
	        {\sin \left(\frac{ \pi}{2}\left(\mu -\frac{1}{2}\right)\right)\Gamma (2 \mu)}\,
	 \Im P_{\mu -\frac{1}{2}}\!\left(\frac{i x}{\delta}\right)
		, \label{eq:smalloddReg} \\
	\xi^{\rm Fr}_{+}(x|\delta) & \equiv &
	\frac{(2\delta)^{\mu-\frac{1}{2}}\Gamma \left(\mu +\frac{1}{2}\right)^2}
	        {\cos \left(\frac{\pi}{2}  \left(\mu-\frac{1}{2}\right)\right)\Gamma (2 \mu)}\,
	 \Re P_{\mu-\frac{1}{2}}\!\left(\frac{i x}{\delta}\right)
		. \label{eq:smallevenReg} 
\end{eqnarray}
Figure \ref{fig:OddEven} shows examples of the unregularized solutions defined in \Eqn{unregParity} and the regularized solutions defined in \Eqn{smalloddReg} and \Eqn{smallevenReg}. 

The leading terms of the asymptotic expansions of these two functions in powers of $\delta^{2\mu}$ are
\begin{eqnarray}
	\xi^{\rm Fr}_{-}(x|\delta) & = & |x|^{\mu-\frac{1}{2}}\,\sgn x
	 +\delta^{2\mu} c_{-}(\mu)\,
	   	|x|^{-\mu -\frac{1}{2}}\,\sgn x + O(\delta^{4 \mu})
		, \nonumber \\
	\xi^{\rm Fr}_{+}(x|\delta) & = & |x|^{\mu -\frac{1}{2}}
	 +\delta^{2\mu} c_{+}(\mu)\,
	   	|x|^{-\mu -\frac{1}{2}} + O(\delta^{4 \mu})
		, \label{eq:smallParityRegAsymp} 
\end{eqnarray}
where the factors $c_{\pm}(\mu)$ are defined by
\begin{eqnarray}
	c_{-}(\mu) & \equiv &
	-2^{2\mu}
	\frac{\Gamma \left(\mu+\frac{1}{2}\right)^2 \Gamma (-2\mu)}
	   	{\Gamma\left(\frac{1}{2}-\mu \right)^2 \Gamma (2 \mu)}
	\frac	{\sin\left(\frac{\pi}{2}  \left(\mu+\frac{1}{2}\right) \right)}
	   	{\sin\left(\frac{\pi}{2}  \left(\mu-\frac{1}{2}\right) \right)}
		 \nonumber\\
	c_{-}(\mu) & \equiv &
	2^{2\mu}
	\frac{\Gamma \left(\mu+\frac{1}{2}\right)^2 \Gamma (-2\mu)}
	   	{\Gamma\left(\frac{1}{2}-\mu \right)^2 \Gamma (2 \mu)}
	\frac	{\cos \left(\frac{\pi}{2} \left(\mu+\frac{1}{2}\right) \right)}
		{\cos \left(\frac{\pi}{2} \left(\mu -\frac{1}{2}\right) \right)}.
		\label{eq:cdefs}
\end{eqnarray}
The $O(1)$ terms are the odd and even generalized function small solutions as required. The $O(\delta^{2\mu})$ terms involve the large solutions of the same parity, with coefficients that vanish as $\delta \rightarrow +0$ because our regularization is based on ideal MHD.

The special case $\mu \rightarrow 1/2$ is important because it corresponds to the zero-$\beta$ limit, or flattening pressure at a rational surface at arbitrary $\beta$. In this limit we find
\begin{eqnarray}
	\xi^{\rm Fr}_{-}(x|\delta) & = & \sgn x - \frac{2\delta}{\pi}\, x^{-1} + O(\delta^2)
		, \label{eq:smalloddRegmuhalf} \nonumber \\
	\xi^{\rm Fr}_{+}(x|\delta) & = & 1 + O(\delta^2)
		. \label{eq:smallParityRegmuhalf} 
\end{eqnarray}

\begin{figure}[htbp]
	\centering
	\begin{tabular}{cc}
	\includegraphics[width=7cm]{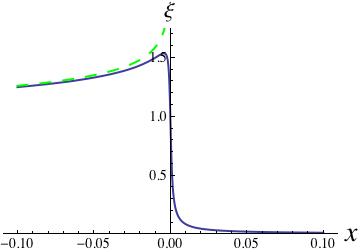}   &  \includegraphics[width=7cm]{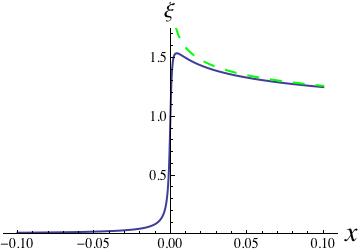}
	\end{tabular}
	\caption{Left panel:  regularized left solution $\xi^{\rm Fr}_{\rm L}(x|0.001)$ (solid line) and  unregularized left solution $\xi^{\rm Fr}_{\rm L}(x)$ (dashed line). Right panel: regularized right solution $\xi^{\rm Fr}_{\rm R}(x| 0.001)$ (solid line) and unregularized right solution $\xi^{\rm Fr}_{\rm R}(x)$ (dashed line). In both cases $\mu = 0.4$.}
	\label{fig:LeftRight}
\end{figure}

We are now ready to regularize Newcomb's ``disconnected'' solutions by combining the regularized odd and even parity solutions in an analogous way to that used for unregularized solutions in \Eqn{unregLR},
\begin{eqnarray}
	\xi^{\rm Fr}_{\rm L}(x|\delta) & \equiv &
	\frac{1}{2}\left[\xi^{\rm Fr}_{+}(x|\delta) - \xi^{\rm Fr}_{-}(x|\delta) \right]
		, \label{eq:smallLReg} \\
	\xi^{\rm Fr}_{\rm R}(x|\delta) & \equiv &
	\frac{1}{2}\left[\xi^{\rm Fr}_{+}(x|\delta) + \xi^{\rm Fr}_{-}(x|\delta) \right]
		, \label{eq:smallRReg} 
\end{eqnarray}
so that the $O(\delta^0)$ terms cancel to the right or left of the origin for $\xi^{\rm Fr}_{\rm L}(x|\delta)$ and $\xi^{\rm Fr}_{\rm R}(x|\delta)$, respectively.

Figure \ref{fig:LeftRight} shows examples of the unregularized solutions defined in \Eqn{unregLR} and the regularized solutions defined in \Eqn{smallLReg} and \Eqn{smallRReg}. It is seen that, for any finite $\delta$, the solutions do connect across the singular point at the origin. By using \Eqn{smallParityRegAsymp} in \Eqn{smallLReg} and \Eqn{smallRReg} we see that the L and R solutions decay rapidly (like the large solution) as $x \rightarrow \pm\infty$ with an amplitude that tends to zero as $\delta \rightarrow +0$, so that their support becomes the negative or positive half line in the limit.

\section{Other MHD intersections}\label{sec:OtherMHD}


The most obvious point of intersection between Allan Kaufman's and my (RLD) research interests is in the oscillation centre theory and the theory of wave action, which goes back to an MHD paper \cite{Dewar_70}. This has developed in various ways that other contributors to this Proceedings will no doubt touch on in much more detail. However, sticking to the MHD theme, it could be said that the theory of flux-minimizing surfaces \cite{Dewar_Meiss_92,Dewar_Hudson_Price_94,Hudson_Dewar_97,Hudson_Dewar_98,Hudson_Dewar_99} is a mutant form of oscillation centre theory, and this has close links with our current research on three-dimensional MHD equilibrium theory \cite{Hudson_Hole_Dewar_07,Hole_Hudson_Dewar_07}.

Another intersecton concerns the theory of quantum chaos, a field in which Allan made an important contribution \cite{McDonald_Kaufman_79}, which is better known outside the field of plasma physics than within it. Indeed there has been very little in the plasma physics literature about quantum chaos, perhaps because of the name. However, this field is really about WKB theory for arbitrary waves in the case of non-integrable ray dynamics, and in a number of papers over the past few years we have applied methods from this field to the study of the ideal-MHD spectrum in a stellarator \cite{Dewar_Cuthbert_Ball_01,Dewar_etal_04,Dewar_Kenny_etal_07}.

\section{Conclusion}\label{sec:Conc}

In this brief paper we have revisited a basic problem in the the theory of ideal MHD stability theory that has connections with Allan Kaufman's early research and our current research. In his long career Allan has been a role model and inspiration to many, and particularly to the first author of this paper.

\section*{Acknowledgements}
The first author (RLD) thanks the organizers of KaufmanFest for giving him the opportunity to help celebrate Allan Kaufman's life and contributions. The algebraic manipulations and plots in this paper were done using \emph{Mathematica} \cite{Mathematica6}. Some of the work mentioned was supported by the Australian Research Council (ARC) Discovery Project DP0452728.

\section*{References}

\bibliographystyle{iopart-num}
\bibliography{RLDBibDeskPapers}

\end{document}